\documentclass[10pt]{sig-alternate-05-2015}
\usepackage{authblk}

\usepackage{graphicx}
\usepackage{color}
\usepackage{xcolor}
\usepackage{xspace}

\newcommand\qt{\texttt{QuantifiedCity} \xspace}

\begin{document}

\toappear{Copyright is held by the author/owner(s).}

\title{The Quantified City: Sensing Dynamics in Urban Setting}

\author{
Tahar Zanouda, Noora Al Emadi, Sofiane Abbar, and Jaideep Srivastava\\
Qatar Computing Research Institute\\
Hamed Bin Khalifa University\\
Doha, Qatar\\
\{tzanouda, nalemadi, sabbar, jsrivastava\}@hbku.edu.qa
}

\maketitle

\abstract{
The world is witnessing a period of extreme growth and urbanization; cities in the 21st century became nerve centers creating economic opportunities and cultural values which make cities grow exponentially. With this rapid urban population growth, city infrastructure is facing major problems, from the need to scale urban systems to sustaining the quality of services for citizen at scale. Understanding the dynamics of cities is critical towards informed strategic urban planning.
This paper showcases \qt, a system aimed at understanding the complex dynamics taking place in cities. Often, these dynamics involve humans, services, and infrastructures and are observed in different spaces: physical (IoT-based) sensing and human (social-based) sensing. The main challenges the system strives to address are related to data integration and fusion to enable an effective and semantically relevant data grouping. This is achieved by considering the spatio-temporal space as a blocking function for any data generated in the city.  
Our system consists of three layer for data acquisition, data analysis, and data visualization; each of which embeds a variety of modules to better achieve its purpose (e.g., data crawling, data cleaning, topic modeling, sentiment analysis, named entity recognition, event detection, time series analysis, etc.) End users can browse the dynamics through three main dimensions: location, time, and event. For each dimension, the system renders a set of map-centric widgets that summarize the underlying related dynamics.   
This paper highlights the need for such a holistic platform, identifies the strengths of the "Quantified City" concept, and showcases a working demo through a real-life scenario.
}

 \section{Introduction}
\label{sec:intro}

'We shape our buildings, and afterwards, our buildings shape us.' wrote Winston Churchill []. 

After the revolutionary growth of the modern industry in the late 18th century, cities have became a niche of new opportunities that attracted huge numbers of local and international migrants. In 2015, we were 53.89\% to live in urban areas and this number is projected to reach 70\% by the end of 2050. This fast urbanization has exposed different services and infrastructures in cities to an increasing stress, leading to a plethora of urban challenges related to human mobility (e.g., traffic congestion), public health (e.g, air pollution), neighborhood deprivation (e.g., lack of parks), and economical difficulties (e.g., unemployment). The situation is such that the United Nation has declared managing urban areas as one of the most important development challenges of the 21st century. 

Luckily, the new cities that we are building are big data generators, which unleashed many initiatives to use data-driven methods in order to tackle the aforementioned challenges and understand more generally how cities work and operate~\cite{batty2012smart}. 
A new body of literature related to \textit{Urban Computing} has been enabled by the availability of data at large scales~\cite{zheng2014urban}. 
On one hand, the advent of the Internet of Things \cite{atzori2010internet} made it easier to collect real-time data about the city dynamics including: infrastructure usage, power consumption, environment, health related issues, human behavior, and commutes. 
At the other hand, the high penetration of smart connected devices (e.g., smartphones) and the rapid adoption of social media platforms, has turned humans into mobile and real-time data generators, leading to what is known as social sensing~\cite{Wang:2015}.   
Thus, researchers have studied different types of dynamics by mining spatio-temporal data generated either by users or sensors. Examples include using call data records (CDR) to understand mobility and commute patterns in cities ~\cite{gonzalez2008understanding, grauwin2015towards}, using crowd-sourced taxi GPS data to estimate gas consumption and air pollution~\cite{Shang:2014}, and finally using check-ins data from location-based social networks (LSBNs) to infer traffic congestion~\cite{Wang:2015}. 

Using IoT-based sensing data has long time tradition in building smart city applications such as parking management systems, whereas using social-based sensing data has been extensively used in building life style applications such as recommender engines for restaurants. We aim in this paper to bridge the gap between these two spaces, by designing \qt -- a comprehensive platform that makes it possible to leverage data from different spaces to build smarter applications. \qt aims at quantifying the dynamics involving humans, services and infrastructures, taking place in cities. A good example to showcase the relevance of this combination are road traffic monitoring systems. Indeed traditional systems rely on the deployment of physical sensors on road segments and intersections to count the number of cars passing by. This allows a nice "real-time" visualization of the traffic status in different parts of the city. However, relying solely on physical sensors, makes it impossible to proceed with the diagnosis of the traffic. A red congested segment, can be so for many reasons (e.g., accident, malfunctioning traffic light, weather condition, etc.) that are completely hidden to the system. The social space provides a nice complementary as many users may have posted online in a completely unsolicited way about the event causing the traffic. Thus, combining the physical space with the social space, modulo the deployment of full-fledged intelligence to link data in the two spaces, may results in semantically richer monitoring dashboards.

\qt consists of three layers for data acquisition, data analysis, and data visualization. Its main contribution is the use of the spatio-temporal space as a blocking function to fuse data that is generated in different spaces (e.g., physical, social). This is inspired from the work of Jaro~\cite{jaro:1989} in which he introduced the use of blocking methods for record linkage. The intuition behind our adaptation is to say that two pieces of information are related if they are generated in same spatial location and within the same time interval. The second big contribution of \qt is the use of cross-modal data mining and machine learning techniques in order to better combine features from the different spaces of data in the knowledge discovery processes~\cite{zheng2014urban}.

We describe in the following sections the global architecture of \qt and provide some details about each of its layers. We also include a demonstration of its capabilities in a real-life scenario.

\begin{figure}[ht]
        \includegraphics[width=0.9\textwidth]{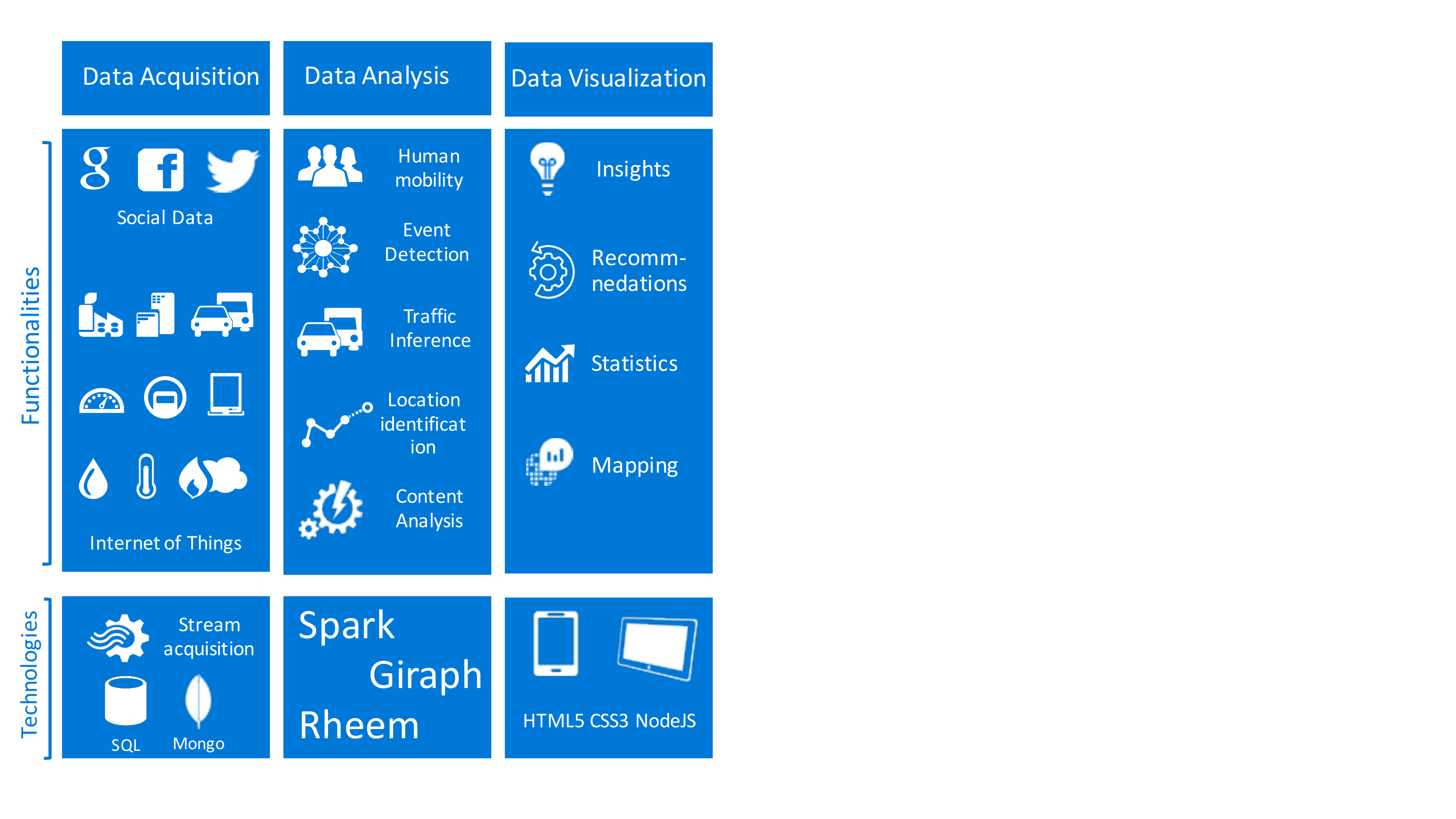}
        \centering
        \caption{Overview of data extraction and data fusion Approach}
        \label{fig:city_multilayer} 
\end{figure}

\section{The Quantified City}

\begin{figure*}[t]
        \includegraphics[width=0.95\textwidth]{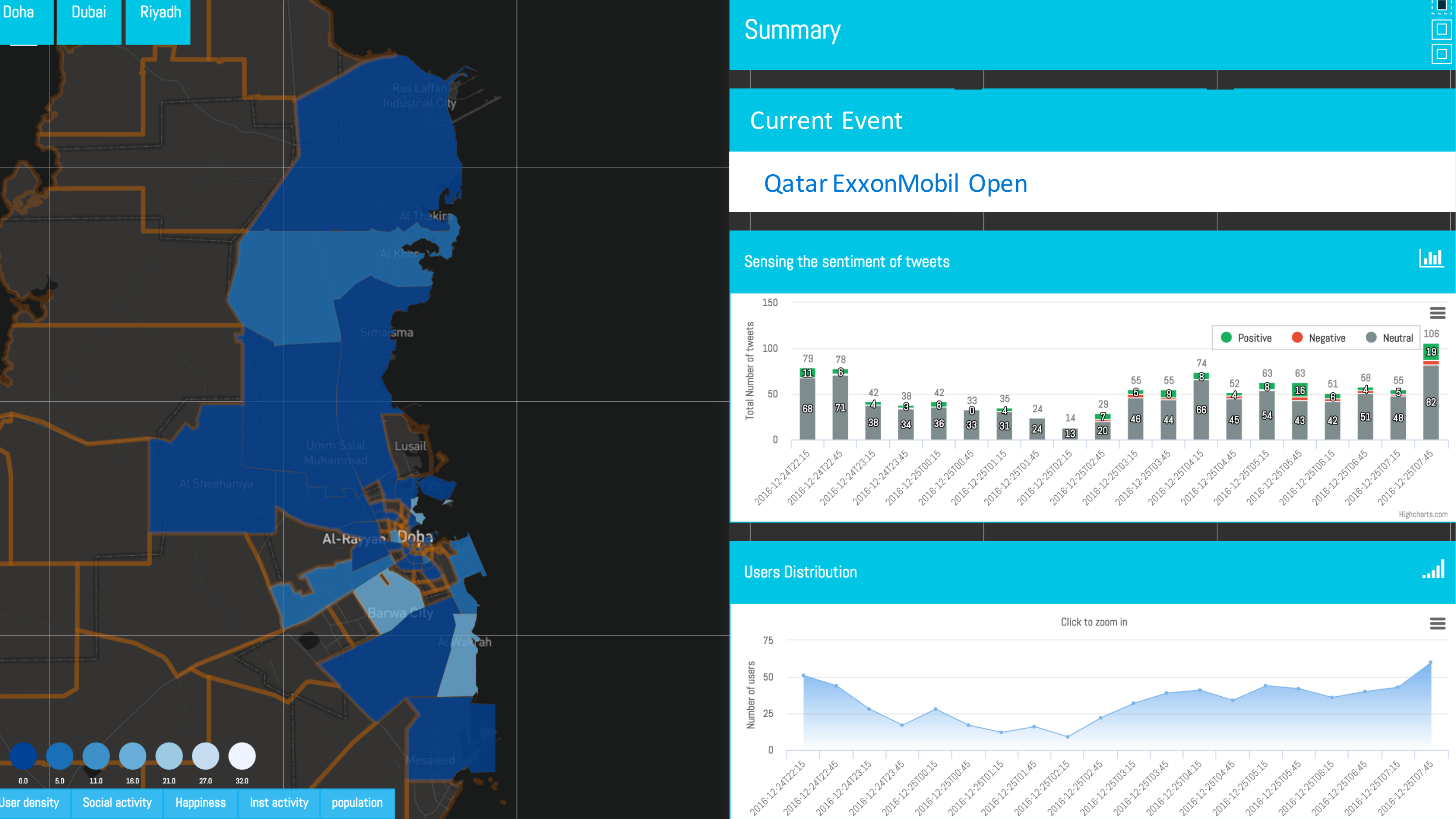}
        \centering
        \caption{Main GUI of \qt. This showcases some statistics around "Qatar ExxonMobil Open" tennis tournament such as overall sentiment per location (map) and time (time line), density of interactions on the map, traffic status, etc.}
        \label{fig:city_pulse_main_gui}
\end{figure*}

The general architecture of the \qt is composed of three major components: (1) data acquisition, (2) data analysis, and (3) data visualization. 
The data acquisition component contains the needed modules to scrap and aggregate streams of data, as well as static geographical data sets such as census. 
Data streams are processed concurrently in real-time and stored in one database. 

The data analysis component contains modules that analyze the data. 
The results of the analysis are visualized through in a map-centric dashboard.  
The system is designed as a suite of independently, small, scalable modules to make it easier and simpler for development.

In what follows, we provide the details of each component.
 
\subsection{Data Acquisition}
This module allows the collection and fusion of several data sets. As a first step, we acquire static geographical and census data \footnote{http://www.mdps.gov.qa/ar/Pages/default.aspx}, this later is used to spatially index the city's data. In order to fuse the knowledge from different data sources, we use spatio-temporal units to conduct our studies. We divide the city into high resolution spatial units, and the day on fine-grained time units to make it easier to study an urban phenomena from different angels. As an example: when we study the effects of on-road traffic on urban air pollution, it's important to collect data on air quality from environmental sensors, number of cars from proximity sensors, and events happening in the same neighborhood from social streams. The cross-referenced datasets can provide insights on relationship between the choice of local events' places and long-term air quality level in the city. This later can lead urban planners to conclusions that can not be identified easily. 

Our system captures, concurrently and continuously, data streams on the city's infrastructure and on the city's social interactions. Once captured, data streams are indexed based on time, and geographical information. Then, data is loaded to centralized geo-database.

\subsection{Data analysis}
This module consists of a set of data mining and machine learning algorithms to enable knowledge discovery from cross-modal fused data. This set of intelligence techniques vary from content-based techniques (sentiment, NER, topic modeling, event detection, etc.), to  structure-based techniques such as network analysis (community detection, etc. ).

As an example, the system is able to capture and mine social content, detect events, and sense the sentiment of people. By mining social networks, and photo-sharing platforms, the system can reveal places that attract people most. Using content-analysis, we can identify places from text, link activity to places and verify whether people complain about traffic using sentiment analysis. Our results can optimize traffic light system in the city to effectively manage big events. 

\subsection{Data visualization}
As we have different mediums of information to showcase, a highly interactive system is needed to allow efficient and smooth interactions with complex interconnected data. The visualization explorer enables users to explore the data easily and intuitively. It is mainly composed of two sub-views, namely, (1) the dashboard, (2) map view.

\texttt{Dashboard View}:  
The dashboard view shows the results of the stats information. The view shows different pages, categorized in main categories and can be accessed by scrolling from top to bottom. Every page has different panels, and every panel contain a chart to visually communicate results of the different analysis.

\texttt{Map view}:
Using Map-centred exploratory approach allows users to easily select spatial choices, as it intuitively provides contextual insights. In addition to the map layer, we use colored zones shapes to showcase the different results. The zones layer enables users to specify the targeted zone.

\section{Demonstration}
\label{sec:demo}

In this section, we showcase a detailed demonstration plan and describe how different users (monitors, urban planners, decision makers, police, etc.) can interact with \qt.

The system provides contextual insights based on three dimensions: i. time, ii. location, and iii. event. As data in our system is indexed to a particular time span, users can select the beginning and the end of the time range, and the granularity of the time analysis (e.g., group data by 20 secs or 5 minutes.) Users are able to narrow focus to a subset of data based on the selected time bounds. Similarly, users can select specific regions or a particular location to filter the data and have limited views to their selections. More interestingly, users can tap into the wealth of data around one event. Subsequently, users can combine different geographical and temporal criteria to gain access to intuitive, yet insightful reports.
 
As an example to illustrate the results, we will focus on some big Tennis tournaments that took place in Qatar this January and show how \qt captures the urban dynamics (social activity, sentiment analysis, fans mobility, traffic congestion, etc.) around these events. With the use of content-analysis methods, \qt can identify and distinguish between events in the same context (i.e., ``Tennis tournament'' will be automatically associated with QATAR EXXONMOBIL OPEN for men during the first week of January, and will be linked to Qatar Total Tournament for women in the last week, thanks to the spatio-temporal blocking function). The system is able to fuse the knowledge from different contextual data to estimate crowd movement patterns (from sensor data about cars moving around the stadiums, and social media activity and engagement), and evaluate the complaints to the city managers during the event (using topic modeling and sentiment analysis). The information can be used to fuel a recommendation engine to help re-routing people from-to the game venues, it can also notify the authorities about the need of an evacuation plan if needed. The system can be connected to the road traffic light system to adjust crowd movement.

Figure \ref{fig:city_pulse_main_gui} shows \qt system visualizing results of Doha city. On the right side of the dashboard, various panels allow users to see the different results of data analysis modules. On the left side, a map with zones polygons are visualized. By clicking on every zone, the system will generate statistics that will take the same form, but projected on one zone.

\section{Conclusion}
\label{sec:conclusion}

In this paper, we presented \qt, a system that captures the complex dynamics taking place in fast growing cities. The system relies on a comprehensive set of data sources coming from different layers: physical overlay through IoT-based sensing and human overlay through social-based sensing which enables a holistic understanding of how cities work and operate. \qt brings a novel perspective into data integration paradigm by using the spatio-temporal space as a blocking function to link parcels of data coming from different sources. The intuition being that data generated in the same geographical space and time window can be grouped together and used for knowledge discovery related to what is happening in that location. 
By bridging the gap between the physical and the social sides of cities, and through an interactive and intuitive map-centric interface, our system allows urban planners and decision makers to visually inspect problems, and interactively analyze and track in real-time the development of events.

\bibliographystyle{abbrv}
\bibliography{biblio} 

\begin{thebibliography}{1}

\bibitem{atzori2010internet}
L.~Atzori, A.~Iera, and G.~Morabito.
\newblock The internet of things: A survey.
\newblock {\em Computer networks}, 54(15):2787--2805, 2010.

\bibitem{batty2012smart}
M.~Batty, K.~W. Axhausen, F.~Giannotti, A.~Pozdnoukhov, A.~Bazzani,
  M.~Wachowicz, G.~Ouzounis, and Y.~Portugali.
\newblock Smart cities of the future.
\newblock {\em The European Physical Journal Special Topics}, 214(1):481--518,
  2012.

\bibitem{gonzalez2008understanding}
M.~C. Gonzalez, C.~A. Hidalgo, and A.-L. Barabasi.
\newblock Understanding individual human mobility patterns.
\newblock {\em Nature}, 453(7196):779--782, 2008.

\bibitem{grauwin2015towards}
S.~Grauwin, S.~Sobolevsky, S.~Moritz, I.~G{\'o}dor, and C.~Ratti.
\newblock Towards a comparative science of cities: Using mobile traffic records
  in new york, london, and hong kong.
\newblock In {\em Computational approaches for urban environments}, pages
  363--387. Springer, 2015.

\bibitem{jaro:1989}
M.~A. Jaro.
\newblock Advances in record-linkage methodology as applied to matching the
  1985 census of tampa, florida.
\newblock {\em Journal of the American Statistical Association},
  84(406):414--420, 1989.

\bibitem{Shang:2014}
J.~Shang, Y.~Zheng, W.~Tong, E.~Chang, and Y.~Yu.
\newblock Inferring gas consumption and pollution emission of vehicles
  throughout a city.
\newblock In {\em Proceedings of the 20th ACM SIGKDD International Conference
  on Knowledge Discovery and Data Mining}, KDD '14, pages 1027--1036, New York,
  NY, USA, 2014. ACM.

\bibitem{Wang:2015}
S.~Wang, L.~He, L.~Stenneth, P.~S. Yu, and Z.~Li.
\newblock Citywide traffic congestion estimation with social media.
\newblock In {\em Proceedings of the 23rd SIGSPATIAL International Conference
  on Advances in Geographic Information Systems}, SIGSPATIAL '15, pages
  34:1--34:10, New York, NY, USA, 2015. ACM.

\bibitem{zheng2014urban}
Y.~Zheng, L.~Capra, O.~Wolfson, and H.~Yang.
\newblock Urban computing: concepts, methodologies, and applications.
\newblock {\em ACM Transactions on Intelligent Systems and Technology (TIST)},
  5(3):38, 2014.

\end{thebibliography}

\end{document}